\documentstyle[prl,multicol,aps]{revtex} 
\input{epsf} 
 
\newcommand{\bq}{\begin{equation}} 
\newcommand{\ee}{\end{equation}} 
\newcommand{\fr}[2]{\frac{#1}{#2}} 
\newcommand{\eps}{\varepsilon}

\begin{document} 
\draft

\title{ Ehrenfest times for classically chaotic systems} 
 
\author{P.G. Silvestrov$^{1,2}$ and C.W.J. Beenakker$^{1}$}

\address{
$^1$Instituut-Lorentz, Universiteit Leiden, P.O. Box 9506, 2300
RA Leiden, The Netherlands\\
$^{2}$Budker Institute of Nuclear Physics, 630090
Novosibirsk, Russia}
\maketitle 
\date{today} 
\begin{abstract} 

We describe the quantum mechanical spreading of a Gaussian wave 
packet by means of the semiclassical WKB approximation of Berry and Balazs.
We find that the time scale $\tau$ on which this approximation breaks 
down in a chaotic system is larger than the Ehrenfest times 
considered previously. In one dimension 
$\tau=\fr{7}{6}\lambda^{-1}\ln(A/\hbar)$, with $\lambda$ the Lyapunov 
exponent and $A$ a typical classical action.

\end{abstract}


\begin{multicols}{2}
%\begin{narrowtext}

According to Ehrenfest's theorem~\cite{Ehrenfest}, the 
propagation of a quantum mechanical wave packet is described for 
short times by classical equations of motion. The time scale at which 
this correspondence between quantum and classical dynamics breaks 
down is called the Ehrenfest time. If the 
classical dynamics is chaotic with Lyapunov exponent $\lambda$, then 
the Ehrenfest time $\tau$ is of order $\lambda^{-1}\ln(A/\hbar)$ 
(with $A$ a typical  
classical action of the dynamical system) \cite{Zaslavsky}. 
There is actually more  
than a single Ehrenfest time, corresponding to different types of  
semiclassical approximations. Although they differ only by a  
numerical coefficient, $\tau_{i}=c_{i}\lambda^{-1}\ln(A/\hbar)$, the  
structure of the wave function changes qualitatively from one time  
scale to the next. 
 
Up to a time
$\tau_{1}$, with $c_{1}=1/6$,
the initial coherent state will retain its Gaussian 
form with vanishing error  
in the limit $\hbar\rightarrow 0$ \cite{Combescure and Robert,Hagedorn}.  
For longer times up to $\tau_2$, with $c_{2}=1/2$,  
the uncertainty in position and momentum 
of the particle remains small
but the phase space structure of the  
wave packet deviates strongly from a Gaussian.
For times greater than $\tau_{2}$  
the wave function no longer has the form of a wave packet 
(this is the ``mixing regime'' of  
Refs. \cite{Bonechi and De Bievre,Bouzouina & Robert}), but up to  
a time $\tau_{3}$ it can still be described semiclassically by the  
time-dependent WKB approximation of Berry and Balazs  
\cite{BerryB}.  
As we will show in this paper, the WKB representation  
implies $c_{3}=7/6$ for a single degree of freedom (with simple  
generalizations for higher dimensions). This is larger than the
value $c_3=2/3$ obtained by Bouzouina and 
Robert~\cite{Bouzouina & Robert} from a different semiclassical 
approximation.

Let us start with the
Gaussian one-dimensional wave packet 
\bq\label{pack} 
\Psi(x)=\left( 
\fr{\alpha}{\pi\hbar} 
\right)^{1/4} 
\exp \left( 
i\fr{p_0 x}{\hbar}+(i\beta-\alpha)\fr{(x-x_0)^2}{2\hbar} 
\right) .  
\ee 
Initially $\beta(t=0)=0$ and $\alpha(t=0)=p_F/L$,  
where $p_F$ and $L$ are the typical classical momentum and length. 
The typical classical action is $A= p_F L$. 
The parameters $x_0(t), p_0(t)$ 
follow the classical trajectory for $\hbar\ll A$.
We will measure momentum and coordinate in units of $p_F$ and $L$, 
respectively, so that $\alpha(0)=1$ and $A=1$.
For chaotic dynamics with Lyapunov exponent $\lambda$ one has 
$\alpha(t)\propto\exp(-2\lambda t)$, hence 
$\alpha\ll 1$ for $t\gg 1/\lambda$.

To describe the time evolution in phase space we 
consider the Wigner function 
\begin{eqnarray}\label{Wig} 
&&W(x,p)=\int\Psi(x+\fr{y}{2})\Psi^*(x-\fr{y}{2}) 
e^{-ipy/\hbar}\fr{dy}{2\pi\hbar}\\ 
&&=\fr{1}{\pi\hbar}\exp\left( 
-\fr{\alpha(x-x_0)^2}{\hbar} 
-\fr{[p-p_0-\beta(x-x_0)]^2}{\alpha\hbar} 
\right) .\nonumber 
\end{eqnarray} 
The wave packet is centered at $x_0(t),p_0(t)$
and for $\alpha(t)\ll 1$ becomes highly elongated and tilted with
slope $\Delta p/\Delta x\approx\beta$.
It has length $l_\parallel=\sqrt{\hbar (1+\beta^2)/\alpha}$ 
and width $l_\perp =\sqrt{\hbar\alpha/ (1+\beta^2)}$, so that  
the area in phase space 
is conserved exactly, 
$l_\parallel l_\perp = \hbar$. 
The Gaussian quantum wave packet satisfies the  
classical Liouville theorem. 

The Gaussian form (\ref{pack}) takes into account the elongation
of the wave packet, but not the curvature that develops in time
and results in a bending of the packet. To describe the curvature 
we add an imaginary 
cubic term in the exponent in Eq.~(\ref{pack}), 
\bq\label{gamma} 
\Psi(x)=\left( 
\fr{\alpha}{\pi\hbar} 
\right)^{1/4} 
\exp \left( 
i\fr{p_0 x}{\hbar}+ 
\fr{(i\beta-\alpha) x^2}{2\hbar} 
+i\fr{\gamma x^3}{3\hbar} 
\right) .
\ee 
(For simplicity we have put $x_0=0$.) 
The cubic term leads to an appreciable phase shift over a length
$l_\parallel\simeq (\hbar/\alpha)^{1/2}$ 
when $(\gamma/\hbar)(\hbar/\alpha)^{3/2}
\gtrsim 1$,
hence when $\alpha(t)\lesssim\hbar^{1/3}\gamma^{2/3}$.

For $\alpha\ll\hbar^{1/3}\gamma^{2/3}$ 
the Wigner function takes again a simple form, 
in terms of the Airy function $\mbox{Ai}$:
\bq\label{airy} 
W(x,p)=
\fr{
\alpha^{{1}/{2}}
\exp(-{\alpha x^2}/{\hbar})}
{\pi\hbar^{{1}/{2}}
(\gamma\hbar^2/4)^{{1}/{3}}}
\mbox{Ai}\left( 
\fr{p_0+\beta x+\gamma x^2-p} 
{(\gamma \hbar^2/4)^{{1}/{3}}} 
\right).
\ee 
One can check that
$W(x,p)\rightarrow\delta(x)\delta(p-p_0)$ when $\hbar\rightarrow 0$
(at fixed $\alpha$), by means of the identity
$\lim_{\eps\rightarrow 0} \mbox{Ai}(z/\eps)/\eps=\sqrt{\pi}\delta(z)$.
At finite $\hbar$ the wave packet is  
extended along the curved line $p=p_0+\beta x+\gamma x^2$. 
Since $p, p_0 ,x$ are of order unity, the two 
parameters $\beta$ and $\gamma$ are of order unity as well
(in contrast to $\alpha$, which is $\ll 1$).
The transverse width
is of order 
\bq\label{scaling} 
l_\perp\approx\gamma^{{1}/{3}} \hbar^{{2}/{3}}
(1+\beta^2)^{-1/2}.
\ee 
The length of the packet remains at $l_\parallel\approx
\sqrt{\hbar (1+\beta^2)/\alpha}$. 
Since now $l_\parallel l_\perp\gg\hbar$, the Liouville theorem
no longer holds.

To obtain the Ehrenfest time we 
parameterize time as 
\bq\label{semilim} 
t=\fr{c}{\lambda}\ln 
\fr{1}{\hbar} 
. 
\ee 
The classical limit for a chaotic system means
$\hbar\rightarrow 0$, $t\rightarrow\infty$ at fixed $c$. 
Different coefficients $c$ follow from different semiclassical
approximations. If we use the Gaussian wave packet (\ref{pack}),
without the cubic term to account for the curvature, then we need 
$\alpha(t)\gg\hbar^{1/3}\gamma^{2/3}$. Since $\alpha\propto 
e^{-2\lambda t}\propto \hbar^{2c}$ we need $c<1/6$.
The upper limit of $c$ gives the first Ehrenfest time 
$\tau_1=\fr{1}{6}\lambda^{-1}\ln(1/\hbar)$.

The classical limit can be reached for longer times if we use 
the wave packet (\ref{gamma}), including the cubic term.
The dimensions of the packet for $t>\tau_1$
scale with $\hbar$ as
\bq\label{scalingc}
l_\perp\propto\hbar^{{2}/{3}} \ , \ l_\parallel\propto   
\hbar^{{1}/{2}-c} .
\ee 
For $c<1/2$ the length of the packet  
approaches zero in the classical limit. 
This upper limit of $c$ gives the second Ehrenfest time
$\tau_2=\fr{1}{2}\lambda^{-1}\ln(1/\hbar)$.

For $t>\tau_2$ 
the length of the wave packet exceeds the size of the system
and is no longer small compared to the 
radius of curvature. For these large times
we may adopt the semiclassical 
WKB approximation of Berry and Balazs~\cite{BerryB}. Consider a
curve in phase space $p(x)$ and a phase space distribution
$\rho(p(x),x)$. Both $p$ and $\rho$ evolve in accordance
with classical equations of motion. 
For $t>\tau_2$ the function $p(x)$ is multivalued with an
exponentially large number of branches 
$\sim \exp[\lambda(t-\tau_2)]$. The quantum wave function 
in this ``mixing'' regime has the form 
\bq\label{semicl} 
\Psi(x)=\sum_k f_k(x)\exp\left(  
{i}\sigma_k(x) /{\hbar}
\right). 
\ee 
The summation over $k$ accounts for the different branches of  
the multivalued function $p(x)$. 
The two functions $f$ and $\sigma$ are related for 
$\hbar\rightarrow 0$ to $p$ and $\rho$ by the correspondence principle,
\bq\label{correspon} 
\fr{d\sigma}{dx}=p(x) \ , \ f=\sqrt{\rho(p,x)} \ . 
\ee 
An explicit description of the evolution 
of the wave function (\ref{semicl}) for quantum maps may be found in  
Ref.~\cite{BerryBalasz}.

Near the point $x_b$ at which $p(x)$ bifurcates into two
branches, one has 
$p=p_b\pm a\sqrt{x-x_b}$, 
$\rho={b}/{\sqrt{x-x_b}}$. 
The wave function there is 
\bq\label{Airy} 
\Psi=\left({\hbar}/{a}\right)^{{1}/{3}} b^{1/2}
\mbox{Ai}\left( 
\left({a}/{\hbar}\right)^{{2}/{3}}(x-x_b)\right) 
e^{ip_b x}, 
\ee 
up to an overall phase. The phase difference between the
bifurcation points can be determined from
Eqs.~(\ref{semicl}) and (\ref{correspon}). 
Because the curve $p(x)$ is not closed, there is no analog 
of the Bohr-Sommerfeld quantization rule.

The Wigner function corresponding to the wave function~(\ref{semicl}), 
being quadratic in $\Psi$, contains both diagonal 
($W_{kk}\propto|f_{k}|^2$) and oscillating nondiagonal 
($W_{km}\propto f_k^\dagger f_m$) contributions. 
Far from 
bifurcations the diagonal contributions to the Wigner function read 
\begin{eqnarray}\label{cartoon} 
&&W_{kk}(x,p)= 
\int\exp\left( 
\fr{iy(\sigma'-p)}{\hbar} 
+ 
\fr{iy^3\sigma'''}{24\hbar} 
\right)
\fr{|f(x)|^2dy}{2\pi \hbar} 
\nonumber 
\\ 
&&=\fr{2}{\sqrt{\pi}}\left( 
\fr{1}{\hbar^2\sigma'''} 
\right)^{{1}/{3}} 
|f(x)|^2 
\mbox{Ai} 
\left( 
\fr{2(\sigma'-p)}{(\hbar^2\sigma''')^{{1}/{3}}} 
\right). 
\end{eqnarray} 
We have made a Taylor expansion of $\sigma(x\pm y/2)$ and neglected 
the difference between $f(x\pm y/2)$ and $f(x)$.

If we parameterize time as in Eq.~(\ref{semilim}) we have for 
both $l_\parallel$ and $l_\perp$ the same scaling with $\hbar$ 
as in Eq.~(\ref{scalingc}). The range of validity of 
Eq.~(\ref{semicl}) is limited by the condition that
different branches 
should be distinguishable. 
This requires that the different parts of the curve $p(x)$  
in phase space should not 
get closer than $\l_\perp$. Their spacing is of order
$1/l_\parallel$ (assuming a uniform filling of phase space), hence 
\bq 
l_\parallel l_\perp\ll 1 \ \Rightarrow \ \hbar^{7/6 -c} \ll 1. 
\ee 
The upper limit of $7/6$ for $c$ leads to the third Ehrenfest time, 
\bq\label{tatau3} 
\tau_3= 
\fr{7}{6\lambda}\ln 
\fr{1}{\hbar} 
. 
\ee 
 
The third derivative $\sigma'''$ in Eq.~(\ref{cartoon}) vanishes  
at the points of inflection of the curve $p(x)$. In order to find the 
Wigner function there one should expand $\sigma(x\pm y/2)$ up to terms 
of order $y^5$. This leads to a different scaling $l_\perp \propto
\hbar^{{4}/{5}}$ of the width of the Wigner function near
the inflection points. Because these are isolated points, they will not
contribute to the matrix elements of nonsingular operators
(containing only smooth functions of $x$ and $p$). This different 
scaling should therefore not affect the Ehrenfest time (\ref{tatau3}).

The nondiagonal contributions $W_{km}$ to the Wigner function 
lead to the ``ghost curves'' discussed in Ref.~\cite{Alonso}.  
The Wigner function near these curves is given by the same Airy  
function as in Eq.~(\ref{cartoon}), but in addition acquires a 
strongly oscillating factor.
Due to these oscillations  
the nondiagonal terms do not contribute to the matrix elements of 
nonsingular operators. 
(They may play a role  in the decoherence by the 
environment~\cite{Zurek}.)
At $t\gtrsim \tau_3$ the ghost curves merge with the  
(multivalued) curve $p(x)$ and become indistinguishable. 
 
The time scale (\ref{tatau3}) for the breakdown of the WKB 
approximation is greater than the Ehrenfest time
$\fr{2}{3}\lambda^{-1}\ln(1/\hbar)$ in the mixed regime 
obtained in Ref.~\cite{Bouzouina & Robert}. That
shorter time scale may signal the breakdown of the series expansion
$\sigma_k(x) \rightarrow \sum_{j=0}\sigma_{kj}(x) \hbar^j$.
Then Eq.~(\ref{correspon}) would no longer hold, but for 
$t<\tau_3$ the representation (\ref{semicl}) with a renormalized
function $\sigma_k(x)$ would still be valid.

So far we have discussed a one-dimensional~(1D) chaotic system, which 
in general can be represented by an area preserving map~\cite{BerryBalasz}.
A familiar example is the kicked rotator~\cite{Chirikov}.
For mesoscopic quantum dots, however, a more 
relevant model is the $d$-dimensional ($d=2,3$) Schr{\"o}dinger equation 
with a smooth potential $V(\vec{r})$. 
The Gaussian wave packet then takes the form 
\bq\label{psismooth} 
\Psi(\vec{r})\propto 
\exp\left[\fr{i}{\hbar}\left(
S(\vec{r}_0(t)) 
+\vec{p}_0\cdot\vec{x}+\fr{\zeta_{ln}}{2}x_lx_n  
\right) \right] 
. 
\ee 
Here $S$ is the action for the classical trajectory
$\vec{r}_0(t)$ and we have defined $\vec{p}_0=m\dot{\vec{r}}_0$,  
$\vec{x}=\vec{r}-\vec{r}_0, 
\zeta_{ln}=\beta_{ln}+i\alpha_{ln}$. 
As before, we rescale momentum and coordinate such that 
the typical classical action $A=1$. 
Initially, ${\zeta}_{ln}\simeq i\delta_{ln} $.  
Similarly to the  one-dimensional case, $\alpha_{ln}$ defines  
the form of the packet in coordinate space 
and $\beta_{ln}=\Delta p_l/\Delta x_n$ give the angles in phase 
space.
Substituting the wave function (\ref{psismooth}) into the  
Schr{\"o}dinger equation one finds Newton's equation  
of motion for $\vec r_0$. The spreading of the wave packet
in phase space is described by
\bq\label{zvec} 
-\dot{\zeta}_{ln}=\fr{1}{m}\zeta_{lk}\zeta_{kn} 
+\fr{\partial^2V}{\partial r_l 
\partial r_n}\bigg{|}_{\vec{r}=\vec{r}_0}. 
\ee 
This is the equation describing the spreading 
in phase space of a small Gaussian bunch of classical
particles.

The Wigner function corresponding to the wave function~(\ref{psismooth}) 
has the Gaussian form  
$W\propto\exp(-Q_l M_{ln}Q_n/\hbar )$, where  
$\vec{Q}=(\vec{r}-\vec{r}_0,\vec{p}-\vec{p}_0)$ is a vector in $2d$  
dimensional phase space. The $d$ Lyapunov exponents $\lambda_i$
($i=1,2,...d$) govern the large-time behavior of the eigenvalues 
$m_i=1/m_{2d-i+1}\propto\exp(2\lambda_i t)$ of the  real symmetric
matrix $M$. Because of energy conservation one Lyapunov exponent
vanishes. We order the $\lambda$'s from large to small,
so that $\lambda_1$ is the largest and $\lambda_d=0$.

The wave packet remains Gaussian (preserving the volume
$\propto \hbar^d$ in phase space) 
until the curvature starts to play a role (via a cubic term in the action).  
The corresponding Ehrenfest time $\tau_1=\fr{1}{6}\lambda_1^{-1}
\ln(1/\hbar)$ is the same as in 1D, 
only now it is defined through  
the largest Lyapunov exponent $\lambda_1$. 
The second Ehrenfest time, when the length of the packet exceeds the 
size of the system, also has the same form $\tau_2=\fr{1}{2}\lambda_1^{-1}
\ln(1/\hbar)$.

The third time $\tau_3$ is different for $d=2,3$ from the 1D case.
Instead of Eq.~(\ref{scalingc}) one now has
\bq
l_\perp^{(i)}\propto\hbar^{2/3} \ , \
l_\parallel^{(i)}\propto\hbar^{1/2}e^{\lambda_i t} \ , \ 
i=1,2,...d-1 \ .
\ee
The longitudinal dimensions $l_\parallel^{(i)}$ correspond to
eigenvalues $m_i$ with $1\le i\le d-1$, and the transverse 
dimensions $l_\perp^{(i)}$ to $m_i$ with $d+2\le i\le 2d$.
The two unit eigenvalues $m_d=m_{d+1}=1$ contribute another factor
$\sqrt{\hbar}$ each to the total volume 
${\cal V}$ in phase space covered by the wave packet:
\bq
{\cal V}=\hbar\prod_{i=1}^{d-1}l_\perp^{(i)}l_\parallel^{(i)}
\propto\hbar^{7d/6-1/6} e^{\lambda_{{\rm tot}}t}\, , \,
\lambda_{{\rm tot}}=\sum_{i=1}^{d-1}\lambda_i  \, .
\ee
The available area ${\cal V}_{\rm max}$ is restricted to a shell of 
constant energy with thickness $\sqrt{\hbar}$, hence
${\cal V}_{\rm max}\propto \sqrt{\hbar}$. We require 
${\cal V}\lesssim{\cal V}_{\rm max}$ for the semiclassical approximation,
which leads to the Ehrenfest time
\bq 
\tau_3=\fr{7d-4}{6\lambda_{{\rm tot}}}\ln 
\fr{A}{\hbar} \, , \, d\ge 2 \, .
\ee

In conclusion, we examined different time scales $\tau_i=c_i\lambda^{-1}
\ln(1/\hbar)$ for the breakdown of different types of semiclassical
approximations. These Ehrenfest times differ only by a numerical
coefficient $c_i$, which may seem insignificant. However, this 
difference is actually a signal of a different power law scaling
with $\hbar$ of the volume ${\cal V}$ in phase space covered by
the wave packet. For short times Liouville's theorem dictates
${\cal V}\propto\hbar$. For long times [parameterized as
$t=(c/\lambda)\ln(1/\hbar)$] 
the WKB approximation gives
${\cal V}\propto\hbar^{7/6-c}$ 
for a one-dimensional
quantum map (such as the kicked rotator) and
${\cal V}\propto\hbar^{7d/6-1/6-c}$ 
for a $d$-dimensional conservative system. These different 
power laws reflect the fundamental change in the structure of the
wave function with increasing time and should therefore have 
observable consequences. Two possible applications are
the Loschmidt echo~\cite{Jalabert} and the quantum shot 
noise~\cite{Larkin}, where the Ehrenfest time plays a key role.

This work was supported by the Dutch Science Foundation NWO/FOM and
by the National Science Foundation under Grant No.~PHY99-07949.

\end{multicols} 
%\end{narrowtext} 
\end{document}